\pdfoutput=1
\documentclass[12pt]{JHEP3}

\let\ifpdf\relax
\usepackage{ifpdf}
\usepackage{graphicx}
\usepackage{amssymb}
\usepackage{amsmath}
\usepackage{slashed}
\usepackage{cite}
\usepackage{epsfig}
\usepackage{datetime}

\newcommand{\be}{\begin{equation}}
\newcommand{\ee}{\end{equation}}
\newcommand{\bal}{\begin{align}}
\newcommand{\eal}{\end{align}}
\newcommand{\bea}{\begin{eqnarray}}
\newcommand{\eea}{\end{eqnarray}}

\newcommand{\tr}{\mathrm{tr}}

\title{Holographic Mutual Information at small
  separations}

\author{Cesar A. Ag\'on  and Howard J. Schnitzer \\
{\small  Martin Fisher School of Physics, Brandeis University, \\ \ \ \ \ \ Waltham, MA 02454, USA\\
}

E-mail: \email{caagon87@brandeis.edu},\\
\email{schnitzr@brandeis.edu}}

\preprint{
 BRX-TH-6291}

\abstract{

The holographic mutual information for the small separation of two circles and two strips in $2+1$ dimensional space-time is considered based on the exact minimal surfaces spanning the boundaries on AdS$_4$. These results provide the leading and sub-leading terms in the short-distance expansion of holographic mutual information. A conjecture for  $d>2$ is also presented, as well as comments about the analogous expansion in conformal field theory.

}

\begin{document}

\section{Introduction}

 Entanglement entropy plays an important role in understanding various aspects of quantum field theory, quantum gravity and black hole physics. Given a reduced density matrix $\rho_A$ for a spatial region $A$, the von Neumann entropy is 
 \begin{eqnarray}
 S_A=-\tr \rho_A \log \rho_A\,
 \end{eqnarray}
one can also consider disjoint spatial regions $A$ and $B$, with mutual information 
\begin{eqnarray}
\label{inf}
I(A,B)=S_A+S_B-S_{A\cup B}\,.
\end{eqnarray}
In (\ref{inf}) area law divergences cancel, while strong subadditivity \cite{Lieb:1973zz,Headrick:2007km} guarantees that $I(A,B)\geq 0$. For example, Cardy \cite{Cardy:2013nua, Schnitzer:2014zva} has considered (\ref{inf}) in the ground state of a $d+1$ dimensional space-time conformal field theory (CFT), in the limit where the separation between $A$ and $B$ is much greater that their sizes $R_A$ and $R_B$. For free scalar field theory, the leading term for the large separation of two spheres is 
 \begin{eqnarray}
 \label{lleading}
 I(A,B)\sim \lambda_d x^{d-1}+ {\cal O}(x^{2(d-1)},x^{d+1})
 \end{eqnarray}
 where 
 \begin{eqnarray}
 \label{xx}
 x=\frac{R_A R_B}{r^2-(R_A-R_B)^2}
 \end{eqnarray}
 is the cross-ratio, with $r$ the separation of the centers of the spheres, and $\lambda_d$ a pure number.
 
 The expansion of $I(A,B)$ for the small separation of two objects is much less developed. In scalar field theory in $d+1$ space-time dimensions, the separation between two objects with parallel faces with areas ${\cal A}_p$ separated by a distance $s\to 0$ has the behavior \cite{Casini:2009sr, Casini:2008wt,Herzog:2014fra}
  \begin{eqnarray}
 \label{lleadingHolo}
 I(A,B)\sim \kappa_d \frac{{\cal A}_p}{s^{d-1}} + \cdots
 \end{eqnarray}
where $\kappa_d$ is a pure number which depends on $d$, and in principle the objects being considered. The corrections to (\ref{lleadingHolo}) are unknown. However, $\kappa_d$ has been computed numerically for two different geometries in \cite{Casini:2009sr, Casini:2008wt} and \cite{Herzog:2014fra}, with agreement between them within numerical errors. This suggests the possibility of a universality for small separation of $A$ and $B$ in $I(A,B)$ in scalar field theory, as is anticipated in \cite{Casini:2009sr, Casini:2008wt}.

One may also consider entanglement entropy when there is a holographic dual to the CFT. Fundamental to this approach is the Ryu-Takayanagi (RT) proposal \cite{Ryu:2006bv,Ryu:2006ef,Nishioka:2009un} for entanglement entropy for a CFT at infinity of anti-de Sitter (AdS) space-time for a subsystem localized in a domain $\cal A $ on the boundary of AdS, where 
\begin{eqnarray}
\label{RT}
S_A=\frac{{\cal A}_A}{4G_N}\,.
\end{eqnarray}
 In (\ref{RT}), $G_N$ is the $d+2$ dimensional Newton constant, $S_A$ the entanglement entropy of $d+1$ Minkowski space-time, with its holographic dual gravitational space-time which is asymptotically AdS and ${\cal A}_A$ is the area of the minimal surface spanning the boundary $\partial A$ defined in a spatial slice of the boundary curve of AdS. The entanglement of several disconnected components can also be considered \cite{Hubeny:2007re}.
 
Results are available for the analytic holographic calculation of $I(A,B)$ for two concentric circles in $d=2$, \cite{Krtous:2013vha, Krtous:2014pva, Fonda:2014cca} and for two strips \cite{Ben-Ami:2014gsa}. After detailed analysis in section \ref{section2} and Appendix \ref{AA}, we find in both cases the holographic mutual information expanded for small s
\begin{eqnarray}
 \label{lleadingHolo2}
 I(A,B)\sim \kappa_d \frac{ {\cal A}_p}{s^{d-1}} + {\cal O}(1)+{\cal O}(s),
 \end{eqnarray}
 where the ${\cal O}(1)$ term is obtained from (\ref{sde}) and (\ref{32}).  We find that 
\begin{eqnarray}
\label{ksp}
(\kappa_2)_{\textrm{circle}}=\frac{R^2}{2 G_N}\frac{[\Gamma(3/4)]^4}{\pi}
\end{eqnarray}
as well as
\begin{eqnarray}
\label{kst}
(\kappa_2)_{\textrm{strip}}=\frac{ R^2}{2 G_N}\frac{[\Gamma(3/4)]^4}{\pi}\,,
\end{eqnarray}
 for two circles and two strips respectively in $d=2$. 
 
 In (\ref{ksp}) and (\ref{kst}), R is the AdS$_4$ radius, and Area ${\cal A}_p$ is the length of the separating curve of the boundary of $\cal A$. The equality of (\ref{ksp}) and (\ref{kst}) suggests a universality of the holographic mutual information as anticipated in \cite{Casini:2009sr, Casini:2008wt} in scalar field theory\footnote{See also \cite{Sabella-Garnier:2014fda} for a similar observation in the mutual information of free scalar fields in commutative and non-commutative geometry}. 
 
 %The existent results also suggests that there may be an analytic procedure for a systematic short-distance expansion of $I(A,B)$, both holographically and in CFT. This is a question that deserves further attention.

\section{Holographic Entanglement of concentric circles \label{section2}}
The analytic expression for the regularized area $\cal{A}$ in AdS$_4$ whose boundaries of two concentric circles are at spatial infinity of the bulk is given in \cite{Krtous:2013vha,Krtous:2014pva,Fonda:2014cca}. This provides a holographic calculation of $I(A,B)$ for two concentric circles $A$ and $B$ in 2+1 dimensional space-time, with results given in terms of elliptic functions which are dependent on $s$, the distance between both circular boundaries of the annulus \footnote{ The disconnected regions whose mutual information is computed is the complement of this annulus}.

We summarize the results of \cite{Krtous:2013vha,Krtous:2014pva,Fonda:2014cca}, emphasizing results we need for a short-distance expansion of $I(A,B)$ in terms of $s$ which requires considerable analysis, as detailed in the text and Appendix (\ref{A}). The AdS$_4$
 metric in Poincar\'e coordinates is 
 \begin{eqnarray}
 \label{Poin}
 \frac{g_{AdS}}{R^2}=\frac{1}{\bar{z}^2}(d\bar{x}^2+d\bar{y}^2+d\bar{z}^2)
 \end{eqnarray}
 where 
\begin{eqnarray}
\label{coor}
&\bar{z}={\bar{r}}/{\cosh \rho}, \quad & \bar{x}=\bar{r} \tanh \rho \cos \varphi \nonumber \\
&\bar{r}=e^{\zeta}\qquad \qquad & \bar{y}=\bar{r}\tanh\rho \sin\varphi\,.
\end{eqnarray}
One presentation of the results for a minimal surface spanned by one circular boundary {\footnote {See \cite{Krtous:2013vha,Krtous:2014pva} for further details}} is
\begin{eqnarray}
{\cal{A}}_{hp}=2\pi R^2(\sqrt{1+P^2}-1)=CR[1-\frac{1}{P}+{\cal{O}}(\frac{1}{P^2})]
\end{eqnarray}
where $C=2\pi R P$ is the circumference of the circular boundary on the cut-off surface $\rho=\rho_*\gg 1$. That is 
\begin{eqnarray}
\label{2.4}
{\cal A}_{hp}=2\pi R^2 P-2\pi R^2+{\cal O}(\frac{1}{P})
\end{eqnarray}
which separates a divergent term from the finite part.

Given two circular boundaries, the boundaries are represented by $\zeta = \pm \zeta_{\infty}$, and $P_0=\cosh \rho_0$. Then, letting $P\to \infty$
\begin{equation}
\label{2.5}
\zeta_\infty(P_0)=\frac{ P_0}{\sqrt{1+P^2_0}\sqrt{1+2P^2_0}}\left[(1+P^2_0)\,{\bf K}\left(\sqrt{ \frac{1+P^2_0}{1+2P^2_0}}\right)-P^2_0\,{\bf \Pi}\left(\frac{1}{1+P^2_0},\sqrt{\frac{1+P^2_0}{1+2P_0^2}}\right)\right]
\end{equation} 
with $s=2R \zeta_\infty (P_0)$. The regularized surface which spans the boundaries of the two concentric circles is
\begin{eqnarray}
\label{2.6}
{\cal A}(P)=2{\cal A}_{hp}+{\cal A}_{ren}+{\cal O}(\frac{1}{P^2})\,.
\end{eqnarray} 
The divergent term ${\cal A}_{hp}$ is given by (\ref{2.4}) and the renormalized ${\cal A}_{ren}$ is 
\begin{eqnarray}
\label{2.7}
\frac{{\cal A}_{ren}}{4\pi R^2}=1+\frac{P^2_0}{\sqrt{1+2P^2_0}}{\bf K}\left(\sqrt{ \frac{1+P^2_0}{1+2P^2_0}}\right)-\sqrt{1+2P_0^2}\,{\bf E}\left(\sqrt{\frac{1+P^2_0}{1+2P_0^2}}\right)\,.
\end{eqnarray} 
Thus, given (\ref{inf}) and the RT conjecture \cite{Ryu:2006bv}, (\ref{2.7}) is proportional to the negative of the holographic mutual information $I(A,B)$ of the two concentric circles.

In this presentation the radius of circles were chosen to be roughly equal to the AdS radius \footnote{  See Appendix \ref{C} for the exact values of the circles's radius }. However, a simply argument\footnote{ From equation (\ref{2.7}), the relation $s=2R \zeta_\infty (P_0)$ and the invariance of the mutual information under overall scaling, it is easy to distinguished between the dependence of the radius of the circle $r$ and of the AdS radius $R$. That is, in (\ref{2.7}) $R$ has to be equal to the AdS radius, otherwise the mutual information would not be scale invariant. From $\zeta_{\infty}=s/2R$ we conclude that $R$ is really the radius of the circles $r$ since $s/r$ is scale invariant and so inverting $P_0=P_0(s/r)$ and inserting into (\ref{2.7}) would guarantee a scale invariant mutual information. } allows us to relax that choice and generalize the previous expressions for arbitrary radius $r$. This only changes the relation $s=2R \zeta_\infty (P_0)$ to $s=2r \zeta_\infty (P_0)$. 

From (\ref{2.5}) and (\ref{2.7}), one has a prediction of $I(A,B)$ as a function of $s$ and $r$, it is known that these equations give $I(A,B)=0$ for $s\ge s_{max}\approx 1.00229 r$. To obtain agreement with the leading large separation for $I(A,B)$ from scalar field theory, one must consider quantum corrections  to (\ref{RT}) in the bulk \cite{Barrella:2013wja, Faulkner:2013ana}.
One can also obtain the short-distance holographic prediction for $I(A,B)$ by expanding (\ref{2.5}) and (\ref{2.7}) valid for small $s<s_{cr}\approx 0.876895 r$. We find the first two terms in the expansion\footnote{ See appendix \ref{AA} for further details}
\begin{eqnarray}
\label{sde}
I(A,B)&\approx &
\frac{4\pi R^2}{4G_N}-\frac{{\cal A}_{ren}}{4G_N}\nonumber \\
&\approx &\frac{(4\pi R^2)}{4 G_N} \left[\frac{\Gamma(3/4)^4}{2\pi^2} \left(\frac{2\pi r}{s}\right)+\frac{\Gamma(1/4)^4}{96\pi^2}\left(\frac{s}{r}\right)+{\cal O}\left(\left({s}/{r}\right)^3\right)\right]
\end{eqnarray}   
We can rewrite this expansion in terms of the conformal invariant cross ratio $x$,  related to the geometric variables by
%$x$,$x\equiv(|p_1-p_2| |q_1-q_2 |)/(|p_1-q_2| |p_2-q_1|)$ where $p_1,p_2,q_1,q_2$ are intersection points of the concentric circles and 
$x=(4 r_+r_-)/s^2=4r^2/s^2$,
\begin{eqnarray}
\label{x}
I(A,B)&\approx &\frac{ R^2}{2 G_N} {\Gamma(3/4)^4}x^{1/2} +\frac{ R^2}{2 G_N}\frac{\Gamma(1/4)^4}{24\pi}x^{-1/2} +{\cal O}(x^{-3/2})\,.
\end{eqnarray}   
This expression gives the leading and sub-leading contributions of the mutual information for geometries conformally related to concentric circles \footnote{See Appendix \ref{D} for the case of disjoint circles.}.

\section{Holographic Entanglement of strips}
Ben-Ami et.al \cite{Ben-Ami:2014gsa} have studied holographic entanglement entropy and mutual information for $m$ strips of varying size and separation. It is interesting to compare their results for $m=2$ strips of equal size with that of Section \ref{section2}. From their equations (2.1) and (2.7) we obtain the expansion for small separations for (2+1) space-time dimensions
\begin{eqnarray}
\label{31}
I(A,B)&=&\frac{4\pi R^2}{4G_N}\left[\frac{\Gamma(3/4)}{\Gamma(1/4)}\right]^2\left(\frac{\bar{L}}{l}\right)\left\{ \frac{l}{s}-\frac 32 -\frac{s}{4l} +\cdots  \right\} \\
\label{32}
&=& \frac {R^2}{2G_N}\frac{\left[\Gamma(3/4)\right]^4}{\pi}\left(\frac{\bar{L}}{l}\right)\left\{ \frac{l}{s}-\frac 32 -\frac{s}{4l} +\cdots  \right\}
\end{eqnarray}
where $R$ is the AdS$_4$ radius, $s$ the separation of the strips, and $\bar{L},l$ are the length and transverse dimension of a trip respectively, with $\bar{L}\gg l$. See Figure 1 of \cite{Ben-Ami:2014gsa}. 

We write the leading term of (\ref{31}) or  (\ref{32}) as 
\begin{eqnarray}
\label{33}
I(A,B)\sim \kappa_2 \frac{ {\cal A}_p}{s}+\cdots 
\end{eqnarray}
From (\ref{32}) we have for (2+1) space-time dimensions
\begin{eqnarray}
\label{34}
(\kappa_2)_{\textrm{strip}}=\frac{R^2}{2G_{N}}\frac{[\Gamma(3/4)]^4}{\pi},
\end{eqnarray}
where ${\cal A}_p=\bar{L}$, is the parallel coincident length of one side of the closed strips. The short-distance expansion for two concentric circles  obtained in section 2 can be similarly expressed as 
\begin{eqnarray}
\label{35}
(\kappa_2)_{\textrm{circle}}=\frac{ R^2}{2 G_N}\frac{[\Gamma(3/4)]^4}{\pi},
\end{eqnarray}
where $R$ is the AdS$_4$ radius, and ${\cal A}_p$=$2\pi r$ is the circumference of a circle that separates the concentric circles with radius $r_\pm$ at the AdS$_4$ boundary \footnote{ See Appendix {\ref{C} for an explanation of this point}}. Comparing (\ref{34}) with (\ref{35}) suggests a possible universality of the leading term in the short distance expansion of holographic mutual information, which is compatible with a similar universality anticipated for CFT in \cite{Casini:2009sr, Casini:2008wt}. Further examples in the holographic context will enable one to explore this possibility. 
 
\section{Concluding remarks}
Results of sections 2, 3 and 4 for the small separation expansion limit of holographic $I(A,B)$ in 2+1 space-time dimensions suggests a universality for the leading term in the expansion.
% with a numerical integer factor depending on the geometry  of A and B.
 It is plausible that there is a similar universality for the short-distance expansion of holographic $I(A,B)$ for d+1 space-time with $d>2$. We conjecture, 
%up to an integer overall factor
\begin{eqnarray}
\label{41}
I(A,B)=\frac{2^d \pi^{d/2}R^d}{(d-1)4G_N}\left[\frac{\Gamma((d+1)/2d)}{\Gamma(1/2d)}\right]^d  {\cal A}_p\left\{\frac{1}{s^{d-1}}+\cdots \right\}
\end{eqnarray}
generalizing results of \cite{Ben-Ami:2014gsa}.

A result analogous to (\ref{lleadingHolo}) is obtained from scalar field theory  in \cite{Casini:2009sr,Casini:2008wt,Herzog:2014fra}
\begin{equation}
\label{gf}
I(A,B)\approx \kappa_2 \frac{{\cal A}_p}{s}+\cdots
\end{equation}
where $\kappa_2\approx 3.97\times 10^{-2}$ \cite{Casini:2009sr}, and  $\kappa_2\approx 3.85\times 10^{-2}$ \cite{Herzog:2014fra} in two different configurations, based on numerical calculations. Results for $d>2$ are also known \cite{Casini:2009sr,Casini:2008wt,Herzog:2014fra}, which anticipates a similar universality for CFT.

Although the holographic and CFT mutual information have the same short-distance power behavior, as expressed by (\ref{lleadingHolo2}), there is no reason to expect the overall dimensionless numerical constants $\kappa_d$ to coincide, as they both occur at weak coupling, and do not reflect a AdS$_{d+2}$/CFT strong-weak duality. 

This is evident in comparing (\ref{34}) or (\ref{35}) with (\ref{gf}). For example, in \cite{Casini:2009sr} one has the estimate for CFT $\kappa_d\approx \Gamma(d/2)/16\pi^{d/2}$ which is roughly an order of magnitude smaller than that obtained from the holographic values (\ref{34}) or (\ref{35}) and the conjectured holographic value from (\ref{41}).

In (\ref{x}) and (\ref{32}) we find the sub-leading term in the short-distance expansion of the holographic mutual information, which by contrast with the leading terms do not appear to exhibit a universal behavior. It would be interesting to study the analogous terms in CFT

Clearly other issues relevant to mutual information at small separations deserve further study \footnote{ We became aware of the paper by Nakaguchi and Nishioka \cite{Nakaguchi:2014pha} while the final version of our manuscript was being prepared. Their sections 4 and 5 are relevant to the issues discussed in our paper}. In that context, it would be interesting to find other minimal surfaces relevant to this issue. 

\section*{Acknowledgements}
We thank Matt Headrick for several useful comments. H.J.S. and C. A. are supported in part by the DOE by grant DE-SC0009987. C. A is also supported in part by the National Science Foundation via CAREER Grant No. PHY10-53842 awarded to M. Headrick.

\appendix

\section{Short distance expansion \label{AA}}
\subsection{Mutual Information \label{A}}
From \cite{Krtous:2013vha},\cite{Krtous:2014pva} we know that the distance $s$ goes to zero as $P_0$ goes to infinity, which means that the short distance expansion in $s$ is a large distance expansion in $P_0$ or short distance expansion in $1/P_0\equiv  z$. Therefore, it is convenient to write equation (\ref{2.7}) in terms of $z$,
\begin{eqnarray}
\label{B.1}
\frac{{\cal A}_{ren}}{4\pi R^2}=1+\frac{1}{z\sqrt{z^2+2}}{\bf K}\left(\sqrt{ \frac{z^2+1}{z^2+2}}\right)-\frac{\sqrt{z^2+2}}{z}\,{\bf E}\left(\sqrt{ \frac{z^2+1}{z^2+2}}\right)\,.
\end{eqnarray} 
We use the conventions of Gradshtein and Ryzhik \cite{Gradshtein}. It is clear that the elliptic integrals ${\bf K}(k),{\bf E}(k)$, are naturally functions of the square of its argument ($k^2$). Then, it is more convenient to do the series expansion around $k\approx 1/\sqrt{2}$ in terms of its square using 
\begin{eqnarray}
k^2&=& \frac{z^2+1}{z^2+2}\approx \frac 12 +\frac{z^2}4 +{\cal O}(z^4)\,.
\end{eqnarray}
We need to expand to second order, since the dominant $1/z$ term is absent. Further, since we want the first two terms, we need to go to second order in the expansion, i.e to $z^4$ in powers of $z$.

Therefore:

\begin{eqnarray}
{\bf K}\left(\sqrt{\frac12 +\frac{z^2}4+{\cal O}(z^4)}\right)&\approx& {\bf K}\left(\sqrt{1/2}\right)+\frac{z^2}4 \frac{d{\bf K}(k)}{d(k^2)}\Bigg  |_{k=\sqrt{1/2}} +{\cal O}(z^4)\nonumber \\
&\approx& {\bf K}\left(\sqrt{1/2}\right)+\frac{z^2}4\left[2 {\bf E}\left(\sqrt{1/2}\right)-{\bf K}\left(\sqrt{1/2}\right)\right]
\end{eqnarray}
and

\begin{eqnarray}
{\bf E}\left(\sqrt{\frac12 +\frac{z^2}4+{\cal O}(z^4)}\right)&\approx& {\bf E}\left(\sqrt{1/2}\right)+\frac{z^2}4 \frac{d{\bf E}(k)}{d(k^2)}\Bigg  |_{k=\sqrt{1/2}} +{\cal O}(z^4)\nonumber \\
&\approx& {\bf E}\left(\sqrt{1/2}\right)+\frac{z^2}4\left[{\bf E}\left(\sqrt{1/2}\right)-{\bf K}\left(\sqrt{1/2}\right)\right]
\end{eqnarray}
where we have used the identities 
\begin{eqnarray}
\label{idei}
\frac{d{\bf K}(k)}{d(k^2)}=\frac{1}{2k^2}\left[\frac{{\bf E}(k)}{1-k^2}-{\bf K}(k)\right], \quad {\textrm{and}} \quad \frac{d{\bf E}(k)}{d(k^2)}=\frac{{\bf E}(k)-{\bf K}(k)}{2k^2}\,.
\end{eqnarray}
Inserting these expansions into (\ref{B.1}), as well as the specific values of ${\bf K}\left(\sqrt{1/2}\right)$ and ${\bf E}(\sqrt{1/2})$, we obtain{\footnote{${\bf K}\left(\sqrt{1/2}\right)=\frac{\Gamma(1/4)^2}{4\pi}$ and ${\bf E}\left(\sqrt{1/2}\right)=\frac{\Gamma(1/4)^2}{8\sqrt{\pi}}+\frac{\Gamma(3/4)^2}{2\sqrt{\pi}}$}}:
\begin{eqnarray}
\label{B.6}
\frac{{\cal A}_{ren}}{4\pi R^2}\approx 1-\frac{\Gamma(3/4)^2}{\sqrt{2\pi}}\frac 1z-\left( \frac{\Gamma(1/4)^2}{16\sqrt{2\pi}}+\frac{\Gamma(3/4)^2}{4\sqrt{2\pi}}\right)z+{\cal O}(z^3)
\end{eqnarray} 
Using equation (\ref{A.5}), we can write an expansion in terms of $\zeta_\infty$
\begin{eqnarray}
\label{B.7}
\frac{{\cal A}_{ren}}{4\pi R^2}\approx 1-\frac{\Gamma(3/4)^4}{2\pi}\frac 1{\zeta_\infty}-\frac{\Gamma(1/4)^4}{48\pi^2}\zeta_{\infty}+{\cal O}(\zeta_\infty^3)
\end{eqnarray} 
or
\begin{eqnarray}
\label{B.8}
\frac{{\cal A}_{ren}}{4\pi R^2}\approx 1-\frac{\Gamma(3/4)^4}{\pi} r/s-\frac{\Gamma(1/4)^4}{96\pi^2}s/r+{\cal O}((s/r)^3)
\end{eqnarray}

\subsection{Distance vs parametrical coordinate\label{B}}
In order to write the short distance expansion of the mutual information, we need to find a simpler short distance relation between the actual distance $s$ and the coordinate $P_0$. Since small $s$ corresponds to large $P_0$ it is convenient to express equation (\ref{2.5}) in terms of $z\equiv 1/P_0$. 
Equation (\ref{2.5}) is then
\begin{eqnarray}
\label{A.1}
\zeta_\infty(1/z)=\frac{ 1}{z\sqrt{z^2+1}\sqrt{z^2+2}}\left[(z^2+1)\,{\bf K}\left(\sqrt{ \frac{z^2+1}{z^2+2}}\right)-\,{\bf \Pi}\left(\frac{z^2}{z^2+1},\sqrt{ \frac{z^2+1}{z^2+2}}\right)\right]\nonumber \\
\end{eqnarray} 
The elliptic integral ${\bf \Pi}(a,k)$ is also a function of the $k$ argument square{\footnote{Notice that ${\bf \Pi}(0,k)={\bf E}(k)$ }} ($k^2$), so again we perform the Taylor expansion in terms of the square of this variable, as well as the variable $a$. That means that up to second order in the derivative expansion this function is: 
\begin{eqnarray}
\label{pp}
&&{\bf \Pi}(a_0+\Delta a ,\sqrt{k^2_0+\Delta k^2})={\bf \Pi}(a_0 ,k_0)+\Delta a \partial_a {\bf \Pi}(a_0 ,k_0)+ \Delta k^2 \partial_{k^2} {\bf \Pi}(a_0 ,k_0)\nonumber \\
&&\qquad  \qquad\frac 12 \left[ (\Delta a)^2 \partial^2_a {\bf \Pi}(a_0 ,k_0)+2\Delta a\Delta k^2  \partial_a \partial_{k^2} {\bf \Pi}(a_0 ,k_0)+(\Delta k^2)^2 \partial^2_{k^2}{\bf \Pi}(a_0 ,k_0)\right]+\cdots \nonumber \\
\end{eqnarray} 
In this case we have to go to second order in $z^2$, since the dominant term in $\zeta_\infty\sim 1/z$ is absent. Thus, we need to express ${\bf K}$ and ${\bf \Pi}$ to order $z^4$.  The expansions
\begin{eqnarray}
a&=& \frac{z^2}{z^2+1}\approx {z^2}-z^4 +{\cal O}(z^6), \nonumber \\
k^2&=& \frac{z^2+1}{z^2+2}\approx \frac 12 +\frac{z^2}4-\frac{z^4}8 +{\cal O}(z^6),
\end{eqnarray}
will be considered explicitly as well as the contribution of the second order derivative expansion of (\ref{pp}). The following identities together with (\ref{idei})  are useful for the evaluation of $\zeta_{\infty}$:
\begin{eqnarray}
\partial_a {\bf \Pi}(0,k)&=&\frac{{\bf K}(k)-{\bf E}(k)}{k^2} \nonumber \\
\partial^2_a {\bf \Pi}(0,k)&=&\frac{2}{3k^4}\left[(2+k^2){\bf K}(k)-2(1+k^2){\bf E}(k)\right]
\end{eqnarray}
leading to the final result:
\begin{eqnarray}
\label{A.5}
\zeta_\infty(1/z)=\frac{\Gamma(3/4)^2}{\sqrt{2\pi}}z-\left( \frac{\Gamma(1/4)^2}{48\sqrt{2\pi}}+\frac{\Gamma(3/4)^2}{4\sqrt{2\pi}}\right)z^3+{\cal O}(z^5)\,.
\end{eqnarray} 

\section{Boundary Area \label{C}}
%The holographic entropy for a single ephere in (2+1) dimensions was found by Casini, et al. \cite{Casini:2011kv}. Their equations (3.24) and (3.25) give
%\begin{eqnarray}
%\label{b1}
%S=\frac{2\pi}{\pi^{3/2}}\Gamma(3/2) \frac{\cal A}{\delta}-2\pi, \qquad \textrm{for} \quad \delta\to 0,
%\end{eqnarray}
%where ${\cal A}_1=2\pi R$, and we set $a_3^*=1$ for convenience. 

%On the other hand, we have for a single sphere from (\ref{2.4})
%\begin{eqnarray}
%\label{b2}
%S&=&\frac{A_{hp}}{4 G_N}\nonumber \\
%&=&\frac{2\pi R^2 P}{4G_N}-\frac{2\pi R^2}{4 G_N} + \cdots
%\end{eqnarray}
%with $P\to\infty$.

%The first term in (\ref{b1}) or (\ref{b2}) is regulator dependent, but the second one is universal. Therefore, we require $(R^2/4G_N)=1$ for consistency.
In the right hand side of equation (\ref{lleadingHolo}), we have the parallel area ${\cal A}_p$ that separates the objects A and B in flat space. The holographic formula also involves an area calculation, but in curved AdS space, whose boundary is identified with the flat space of the field theory.
%That means that  a given minimal surface whose boundary section (AdS boundary) is equal to ($\partial \cal A$) has an area area($\partial \cal A$) from the point of view of the field theory  that can be computed from the AdS point of view by going to Poincar\'e coordinates and calculate its area in the Minkowski metric of its boundary.
 For example, in \cite{Krtous:2013vha,Krtous:2014pva} the two circular boundaries are represented by $\zeta=\pm \zeta_{\infty}$. Then, translating this to Poincare coordinates (\ref{Poin}, \ref{coor}), the boundary circles will have radius $r_{\pm}=r e^{\pm \zeta_\infty}$ at $z=0$, where the metric (\ref{Poin}) diverges but the Minkowski boundary metric $(dx^2+dy^2)$ does not, and leads to the boundary circumferences\footnote{We use here non-bar notation for full dimensional coordinates related to $\bar{x}^i$ by $x^i=r\bar{x}^i$}  $A_{\pm}=2\pi r e^{\pm s/2R}$. In the short distance limit $A_+\approx A_-$ and it is a matter of convenience what area we will choose as the parallel area ${\cal A}_p$ as soon as $A_- \leq {\cal A}_p\leq A_+$. We conveniently chose it to be the one given by the geometric average $r_0=\sqrt{r_+ r_-}$ between the inner and otter circle's radius which corresponds to ${\cal A}_p$=$2\pi r$. 
 
\section{Disjoint Circles \label{D}}

The two concentric circles can be mapped by a conformal transformation to two disjoint circles as discussed recently in \cite{Nakaguchi:2014pha}. The conformal ratio $x$, in terms of the new geometric variables is $ x=(4 r'_+r'_-)/[s(2(r'_++r'_-)+s')]$, choosing  $r'_+=r'_- =r'$ leads to $x\approx r'/s'$ which together with (\ref{x}) gives the leading mutual information for disjoint circles\footnote{We thank Matt Headrick for giving us an intuitive understanding of the scaling behavior $\sqrt{r'/s'}$}
\begin{eqnarray}
I(A,B)\approx \frac{R^2}{2 G_N} \Gamma(3/4)^4 \sqrt{\frac{r'}{s'}}\,.
\end{eqnarray}

%\eject
\bibliographystyle{utphys}

\bibliography{emirefs}

\end{document}